# HEROHE Challenge: assessing HER2 status in breast cancer without immunohistochemistry or in situ hybridization

## Authors

Eduardo Conde-Sousa[a,b], João Vale[a,c], Ming Feng[d], Kele Xu[e], Yin Wang[d], Vincenzo Della Mea[f], David La Barbera[f], Ehsan Montahaei[g], Mahdieh Soleymani Baghshah[h], Andreas Turzynski[i], Jacob Gildenblat[j], Eldad Klaiman[k], Yiyu Hong[l], Guilherme Aresta[m,n], Teresa Araújo[m,n], Paulo Aguiar[a,b], Catarina Eloy[a,c,o], António Polónia[a,c]

a. I3S - Instituto de Investigação e Inovação em Saúde, Universidade Do Porto, Portugal
b. INEB - Instituto de Engenharia Biomédica, Universidade Do Porto, Portugal
c. Department of Pathology, Ipatimup Diagnostics, Institute of Molecular Pathology and Immunology, University of Porto
d. Tongji University, Shanghai, China
e. National University of Defense Technology, Changsha, China
f. Department of Mathematics, Computer Science and Physics, University of Udine, Udine, Italy
g. Sharif University of Technology, Tehran, Iran
h. Computer Engineering Department, Sharif University of Technology, Tehran, Iran
i. Private Group Practice for Pathology, Lübeck, Germany
j. DeePathology, Hatidhar 5, Raanana, Israel
k. Roche Diagnostics GmbH, Nonnenwald 2, 82377 Penzberg / Germany
l. Department of R&D Center, Arontier Co., Ltd, Seoul, Republic of Korea
m. INESC TEC - Institute for Systems and Computer Engineering, Technology and Science, Porto, Portugal
n. FEUP - Faculty of Engineering, University of Porto, Porto, Portugal
o. FMUP - Faculty of Medicine, University of Porto, Porto, Portugal

**Corresponding author**

Eduardo Conde-Sousa (econdesousa@gmail.com)

INEB/i3S, Rua Alfredo Allen, 208; 4200-135 Porto, Portugal; Tel: +351 226074975

## Highlights

- The HEROHE Challenge was organized to promote the development of computer-aided diagnosis tools to assess human epidermal growth factor receptor 2 status in invasive breast cancer samples stained with hematoxylin and eosin.
- Two whole slide images datasets of invasive breast cancer tissue samples were compiled to be used as training (359 cases) and test (150 cases) datasets.
- A total of 21 methods were developed, submitted, and ranked according to their performance on the test dataset. A comparative analysis was performed and the top-6 ranked for the $F_1$ score are described.
- The large majority of the methods rely on pre-trained convolution neural networks. More than half of these methods (including 5 of the top-6) result from the combination of more than one network.

## Graphical Abstract

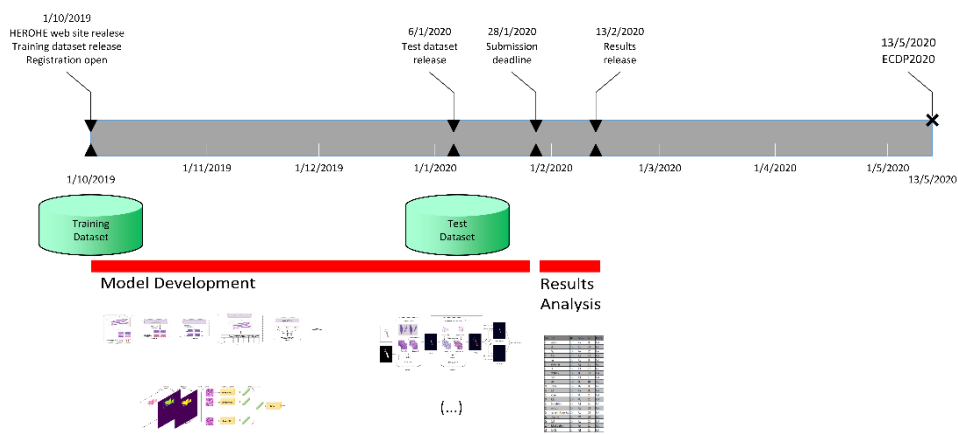

## Abstract
Breast cancer is the most common malignancy in women, being responsible for more than half a million deaths every year. As such, early and accurate diagnosis is of paramount importance. Human expertise is required to diagnose and correctly classify breast cancer and define appropriate therapy, which depends on the evaluation of the expression of different biomarkers such as the transmembrane protein receptor HER2. This evaluation requires several steps, including special techniques such as immunohistochemistry or in situ hybridization to assess HER2 status. With the goal of reducing the number of steps and human bias in diagnosis, the HEROHE Challenge was organized, as a parallel event of the 16[th] European Congress on Digital Pathology, aiming to automate the assessment of the HER2 status based only on hematoxylin and eosin stained tissue sample of invasive breast cancer. Methods to assess HER2 status were presented by 21 teams worldwide and the results achieved by some of the proposed methods open potential perspectives to advance the state-of-the-art.

## 1. Introduction
Breast cancer (BC) is the most common cancer among women. More than two million women are diagnosed each year with BC, being this disease the cause of more than half a million deaths every year, representing roughly 15% of all cancer-related women deaths [1].

### 1.1. Breast cancer diagnosis
BC detection usually starts with self-check-ups via palpation and regular screenings through imaging techniques (ultrasounds and/or mammography). When an abnormality is detected further examinations can be performed. A breast biopsy consists of the sampling of breast tissue through a needle, processed and stained with hematoxylin and eosin (H&E) allowing the visual observation under the optical microscope. A pathologist observes the sample and assesses the existence of cancer signs, identifying malignant and non-malignant tissue regions. The malignant regions are then classified as *in situ* carcinoma when restrained to the mammary ductal-lobular system, or invasive if cancer cells are spread outside the ducts [2].

## 1.2. HER2 assessment

Current guidelines [3, 4] recommend routine quantification of ER (estrogen receptor), PgR (progesterone receptor), and HER2 (human epidermal growth factor receptor 2) status in all patients with invasive BC, recurrences, and metastases. HER2 is a transmembrane protein receptor with tyrosine kinase activity, being amplified and/or overexpressed in approximately 15% of BC [4]. These BC cases are classified as HER2-positive, being associated with aggressive clinical behavior but also with better responses to HER2-targeted therapies. Several clinical trials have shown an association between these therapies and a significant improvement in disease-free survival and overall survival for patients with HER2 positivity [5-9], thus making of paramount importance the correct identification of this BC subtype.

Usually, HER2 evaluation begins with the analysis of protein expression using specific antibodies that recognize the protein by immunohistochemistry (IHC). In this test, the following results can be achieved: negative (score 0 or 1+), equivocal (score 2+), positive (score 3+), and indeterminate. Equivocal and indeterminate samples require a reflex test, consisting of the evaluation of HER2 amplification with either fluorescence or bright field in situ hybridization (ISH) assays [4, 10]. IHC is easier to perform than ISH, although the later test is more robust but also more expensive [10, 11] and can ultimately classify BC samples in HER2-positive and HER2-negative providing the basis for the application of HER2-targeted therapy.

Like most ancillary tests in pathology laboratories, both IHC and ISH tests are sensitive to pre-analytical conditions, such as ischemic time, type of fixative, and duration of fixation [4]. The above-mentioned conditions can compromise the results of the tests, being responsible for the presence of false-negative and false-positive results, which can constitute a major impact on the effectiveness of the implemented treatment.

## 1.3. Digital pathology

The approval of digital pathology (DP) systems by the US Food and Drug Administration (FDA) has accelerated the implementation of DP in many pathology departments across the globe [12-17]. There are several advantages described in the literature for using whole-slide images (WSI) instead of glass slides. These include instant sharing of slides for educational purposes or internal/external consultation of challenging cases as well as for the practice of telepathology [18]. Nevertheless, the main advantage of WSI is probably the potential of application of image analysis tools for *in silico* evaluation that could go beyond traditional quantification analysis, such as IHC analysis, and achieve qualitative analysis to create computer-aided diagnosis (CAD) tools.

## 1.4. Computer-aided diagnosis

By relying on the visual assessment of the pathologist, the described methodology is prone to human bias and is time-consuming, thus several studies have been implemented aiming at standardizing methodologies and developing automatic methods to BC diagnosis. In previous works [19, 20] methods for automatic nuclei segmentation and feature extraction were presented allowing the application of different classifiers to differentiate between benign and malign BC.

Other studies try to solve the more complex 3-class problem, discriminating between normal tissue, in situ carcinoma, and invasive carcinoma [21]. In [22], the 4-class problem (classifying breast tissue as normal tissue, benign, in situ carcinoma, and invasive carcinoma) was tackled by manually extracting features while in [23] a deep learning approach was taken.

The described methods rely on image data to classify the tissue in 2, 3, or 4 classes but none tackles the subsequent steps for HER2 status assessment in invasive BC. This problem was previously addressed in [24] where the HER2 and hematoxylin stainings were separated through color deconvolution to allow the individual cell classification into seven classes (immune, stroma, artifacts, T0, T1+, T2+, and T3+).

In [25], a standard machine learning approach was used to quantify IHC breast cancer images. The prediction of molecular subtypes in BC was attempted using image analysis of H&E and deep learning methods [26]. The work consisted of the evaluation of BC histological images from the Carolina Breast Cancer Study in a tissue microarray (TMA) with molecular classification performed using the PAM50 gene signature. The authors were able to correctly classify high-grade tumors and ER status with accuracies above 80%. The molecular classification accuracy was less impressive (77%) and it was not able to classify in the usual 4 subgroups but only in two larger subgroups (basal versus non-basal subtypes). One of the limitations of this study was the use of TMAs that, in this case, consisted of just 1 to 4 tumor tissue cores per patient with 1mm of diameter, which might be an insufficient amount of tumor to be analyzed and might have compromised the extraction of features for subsequent image classification. Nevertheless, the work provided a strong proof of principle that molecular classification can be predicted based on the extraction of H&E features.

The prediction of the expression of molecular biomarkers in breast cancer based only on the evaluation of digitized H&E-stained specimens was also attempted by Shamai and colleagues [27]. In this work, a deep convolutional neural network (CNN) based on a residual network (ResNet [28]) architecture was developed to predict 19 biomarkers, including ER, PgR, and HER2 from tissue morphology. For these 3 biomarkers, the area under the receiver operating characteristic curve (AUC) was 0.80, 0.75, and 0.74, respectively. The data originated from a single institution (Vancouver General Hospital) and included only TMA images from 5356 patients, rather than WSI, representing two important limitations. Ultimately, the system was able to predict biomarker expression with noninferiority to traditional IHC assay.

More recently, Naik and colleagues [29] developed a multiple instance learning-based deep neural network to predict the same molecular biomarkers from H&E-stained WSI. The algorithm based on a ResNet50 was trained in a multi-country dataset of 3474 patients (Australian Breast Cancer Tissue Bank and The Cancer Genome Atlas), and achieved an AUC of 0.92, 0.81, and 0.78, for ER, PgR, and HER2, respectively.

Following the same rationale, Kather and co-workers [30] developed a deep learning model based on ShuffleNet [31] to predict molecular alterations in 14 of the most common solid tumor types, including breast cancer. The system trained on The Cancer Genome Atlas dataset was able to infer from histology images alone at least one mutation from all except one tested tumor type. In breast cancer, ER, PgR and HER2 subtype could be predicted with an AUC of 0.82, 0.74, and 0.75, respectively.

### 1.5. International challenges

Challenges are excellent opportunities to advance the state-of-the-art in any given field by gathering experts with different backgrounds to solve one scientific question and thus promoting a proper balance between competition and collaboration. Over the past years, several challenges were open to the community to solve problems related to Breast Cancer.

The MITOS-ATYPIA-14 challenge (https://mitos-atypia-14.grand-challenge.org) was developed in late 2013 aiming to detect mitosis and evaluate of nuclear atypia score on BC histological images stained with H&E.

In 2015, the BIOIMAGING2015 challenge [23] asked participants to develop automatic methods to classify H&E stained breast biopsy tissue samples in four classes: 1) normal, 2) benign, 3) in situ carcinoma and 4) invasive carcinoma.

In 2016, the HER2 Scoring Contest [32] was proposed to compare and advance the state-of-the-art artificial intelligence-based methods to automate HER2 scoring. The challenge consisted in the development of a scoring method for whole slide images (WSI) stained with H&E and IHC for HER2.

In 2016 and 2017, two consecutive editions of the CAMELYON challenge [33, 34] were open to detect breast cancer metastases in H&E stained WSI of lymph node sections.

In 2018 the BACH challenge [35] was released divided into two parts. The first part was a follow-up of the BIOIMAGING2015 while the second part consisted in the development of a semantic segmentation method, where each WSI should be pixel-wise labeled into one of the four classes.

More recently, in late 2018, the BreastPathQ Cancer Cellularity Challenge (https://breastpathq.grand-challenge.org/) focused on the development of automated methods for analyzing tumor-burden assessment in breast pathology using histology patches extracted from WSI.

A deeper analysis of these and other public challenges to solve different pathology problems through the application of artificial intelligence can be found in [36].

### 1.6. HEROHE Challenge

The HEROHE (HER2 On H&E) Challenge was developed aiming to predict the HER2 status in invasive BC samples by the analysis of H&E slides, without access to the IHC or ISH assays. Image analysis algorithms for HER2 prediction may not only decrease considerable costs for pathology laboratories but also serve as safety nets for the usual analysis of HER2 by IHC and ISH. The HEROHE Challenge aimed to promote the production of image analysis algorithms able to, at least, replace a considerable amount of HER2 tests in BC. This would allow reducing the costs of pathology exams and accelerate HER2 status determination, and/or pinpoint cases that despite being conclusive by IHC could benefit from additional testing to reduce the existence of false-results. This tool could also be used to select the sample most likely to be positive in the case of patients with multiple samples, reducing the cost of analyzing all the samples.

Although pathologists rely on IHC and/or ISH assays for the evaluation of HER2 in BC, previous literature shows that HER2-positive BC is associated with different morphological features compared to HER2-negative BC. These features consist of poor differentiation (more solid tumors without tubule formation), higher nuclear pleomorphisms (high nuclear grade), and higher number of mitosis, which are all aggregated in the establishment of histological grade [37-39]. Although other features can probably differentiate between these two molecular subtypes of BC, some of which pathologists may have never realized, this establishes the morphological basis for the success of the proposed task. In addition, previous deep learning models have been used to predict IHC images from the H&E slides [40-42] thus establishing the computational basis for the success of the HEROHE challenge.

In this work, we overview the organization steps of the HEROHE challenge, the first developed to predict HER2 status from H&E stained WSI, and present the methods and results obtained by participant teams.

## 2. Material and Methods

The HEROHE Challenge was organized as a parallel event of the 16th European Congress on Digital Pathology (ECDP2020). Although the ECDP2020 was canceled due to the coronavirus pandemic, the HEROHE Challenge was performed successfully.

### 2.1. HEROHE challenge organization

The HEROHE Challenge website was hosted on the Grand Challenge servers with the domain https://ecdp2020.grand-challenge.org/. The Grand Challenge is one the largest platforms on medical imaging challenges, counting, at the time of writing, with more than 40,000 users. Hosting HEROHE at the Grand Challenge website allowed for an easy setup while maximizing the number of researchers reached. The challenge was also advertised through the social media networks and official webpage of the ECDP2020, and monetary prizes were awarded to the three best-performing methods.

Unlike previous challenges, where IHC images were part of the training and testing datasets, the goal here was to classify HER2 status directly from the morphological features present on the H&E stained images. Thus, the training dataset consisted of 359 WSI of invasive BC tissue samples stained only with H&E and the corresponding image-wise ground truth classification based on IHC and ISH. The cases did not include annotations such as the location of the invasive carcinoma and no IHC or ISH slides were provided. The ground truth originated from IHC and ISH tests, resulting in a binary classification (negative or positive). Considering the IHC HER2 scores, the dataset had 43 (12%) cases scored of 0, 47 (13%) scored of 1+, 230 (64%) score of 2+, and 39 (11%) scored of 3+. One case with a score of 1+ had HER2 amplification by ISH, being classified as HER2 positive. In cases with a score of 2+, 126 cases were ISH-negative and 104 ISH-positive. Hence, the dataset contained 144 HER2 positive cases (40%) and 215 HER2 negative cases (60%). There were 358 female and 1 male cases, with ages between 24 and 92 (median 58 years old). Cases originated in 22 different laboratories and all ISH tests were performed at Ipatimup Diagnostics (national reference center for HER2). Cases with HER2 heterogeneity were not included in the dataset. All cases were classified by two experienced pathologists (CE and AP) according to the latest American Society of Clinical Oncology/College of American Pathologists (ASCO/CAP) guidelines for BC, scanned at Ipatimup Diagnostics in a 3D Histech Pannoramic 1000 digital scanner at 20x magnification and saved in the MIRAX file format.

On October 1st, 2019, the HEROHE Challenge website and training dataset were released.

The test dataset was released on January 6th, 2020. In total 150 WSI correspondent to 150 cases were acquired following the same conditions of the training dataset, including the proportion of positive and negative cases (test dataset distribution was not previously known by the participants). Among the 150 WSI of the test dataset, 19 (13%) with a score of 0 on the IHC test, 18 (12%) with a score of 1+, 85 (57%) with a score of 2+, 27 (18%) with a score of 3+ and 1 case without IHC score. In cases with a score of 2+, 53 cases were ISH-negative and 32 ISH-positive. The case without IHC score was HER2-positive by ISH. Hence, the dataset contained 60 HER2 positive cases (40%) and 90 HER2 negative cases (60%). There were 149 female and 1 male cases, with ages between 33 and 93 (median of 57 years old) from 17 different pathology laboratories. All cases from test and training datasets originate from different patients to ensure independence between datasets.

We were also able to trace 116 cases, from both the training and the test datasets, that showed positive (score of 3+) or negative (score of 0 or 1+) results by IHC and the corresponding ISH results obtained either by an internal or external quality control protocol. In these cases, there was only one false-negative case by IHC (mentioned above), providing a sensitivity of 0.98, a specificity of 1.00, a positive predictive value of 1.00, and a negative predictive value of 0.98 for the IHC analysis.

To participate and be eligible for the Challenge's prizes at least one element of each competing team should be registered to ECDP2020 and submitted, until January 28[th], 2020, the methods code, the test dataset prediction (hard and soft predictions), and a short method description. The ECDP2020 registration requirement was later removed from the challenge rules due to the cancellation of the congress.

## 2.2. Evaluation

For the ranking of the proposed methods, the $F_1$ score, the harmonic mean between precision and recall, was used:

$$F_1 = \frac{2}{P^{-1} + R^{-1}} = 2 \times \frac{P \cdot R}{R + P} \qquad [1]$$

where $P = \frac{tp}{tp+fp}$ is the precision, $R = \frac{tp}{tp+fn}$ is the recall, $tp$, true positives, is the number of positive cases classified as positive, $fp$, false positives, is the number of negative cases classified as positive, and $fn$, false negatives, is the number of positive cases classified as negative. Besides the $F_1$ score, other metrics were also assessed, namely the area under the curve (AUC), precision, and recall, although not being considered for the ranking of each competing team. The receiver operating characteristic (ROC) curve is a graphical plot of the true positive rate (TPR), also known as recall, against the false positive rate ($FPR$), at various threshold values. $FPR = \frac{fp}{fp+tn}$, where $tn$, true negatives, is the number of negative cases classified as negative. Since the ROC curve is a two-dimensional curve, to compare methods the entire curve should be collapsed in one single real number, and the most common method is to calculate the AUC [43].

## 2.3. Competing Solutions

A total of 21 teams or individual participants submitted their methods until the challenge deadline. Bellow, we briefly describe the methods proposed by the six best-ranked teams. A brief description of the remaining methods is presented in the supplementary material.

### 2.3.1. Team Macaroon

Team Macaroon employed a two-stage method to solve the problem. In stage A, a ResNet34 model, pre-trained on the CAMELYON16 [34] challenge datasets, was used for training a patch-based (256 by 256 pixels) classification model to differentiate normal tissue patches from tumor patches (see [44] for training details). A probability map, PM_A, results from each WSI. In stage B, each original WSI was down-sampled (ratio 1:2) and a sliding window split it into 256 by 256 pixels patches to be classified by the model from stage A. Potential tumor patches were extracted and the HER2 status information of the WSI was added to each patch. The resulting dataset was used for training a second ResNet34 model aiming to classify tumor patches as HER2 positive or HER2 negative. Model B was used for generating the final probability map of the WSI (denoted as PM_B). After training, a new WSI is classified as HER2 positive if more than 50% of the tumor patches (those where PM_A > 0.5) are classified as positive by the network B (PM_B > 0.5), and is classified as negative otherwise. The overall architecture of the resulting model is in Figure 1.

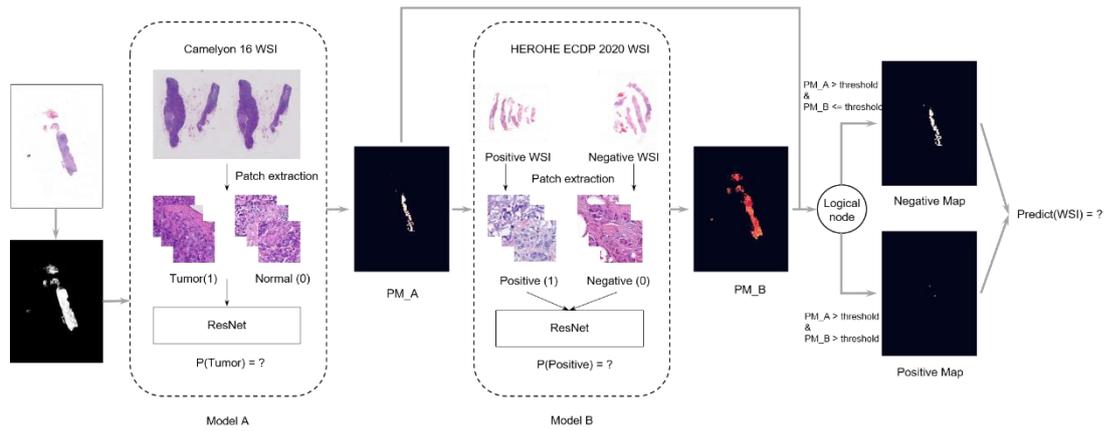

Figure 1: Overall architecture of the model developed by team Macaroon

### 2.3.2. Team MITEL

A five-stage procedure was used by team MITEL (see Figure 2). The full method is described in [45]. A preprocessing step was implemented where each WSI was first downsampled (ratio=1:2) and tiles were created by a sliding window of 512 by 512 pixels. Only tiles with an average grey level of <85% were retained. In the second stage, Tumor Detection, tiles were classified as tumor or normal by a DenseNet201 [46]. The model used was pre-trained on ImageNet [47] and then fine-tuned on the BACH [35] dataset for the tumor classification task. Tiles classified as normal tissue were discarded while the others were used in the following stage. In the third stage, HER2 classification, the remaining tiles were fed to a ResNet152 (optimized for precision), and the model was trained to predict the probability of a given tile being from a HER2 positive WSI. In the fourth stage, results from all tiles of any given WSI were aggregated into three WSI-level features:

1. *Overall positivity*: mean positivity probability for all tiles in a WSI. If above 0.5 then the slide is positive for HER2;
2. *Strength of positivity*: mean positivity probability of positive tiles only. If above 0.66 then the slide is positive;
3. *Extent of positivity*: percentage of positive tiles. If 35% of the tiles for each slide is positive, then the slide is positive.

Finally, in the fifth stage, each WSI is classified as the result of a majority voting among the results of the three conditions.

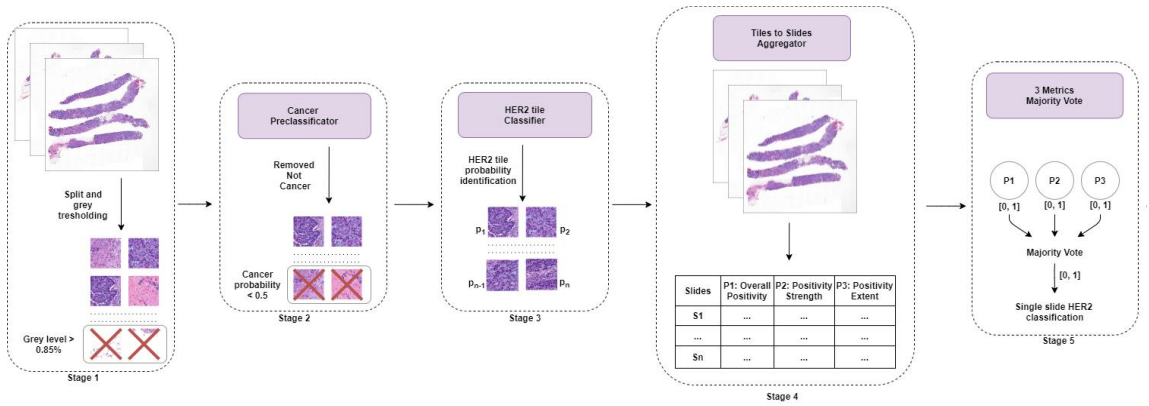

Figure 2: Overall architecture of the model developed by team MITEL

### 2.3.3. Team Piaz

The team Piaz used a four-stage procedure (https://github.com/IAmS4n/HEROHE) to classify each WSI in a multi-instance learning fashion (see Figure 3). In stage one, the Shannon entropy of the WSI was computed to identify the most informative regions, and then a threshold, based on the method of minimum value in histogram [48], was found to make a tissue mask. Next, 256 random patches of 222 by 222 pixels were extracted from the tissue mask's valid regions at a maximum resolution. In the second stage, an EfficientNet-B0 [49], pre-trained on the BACH challenge [35] dataset, was used and retrained on the HEROHE dataset to extract a 64-dimensional feature array for each patch. The features were extracted using a max-pooling over the last CNN layer of EfficientNet. In addition, there was a batch-norm layer followed by an absolute operation after each max-pooling. In the third stage, a novel pooling function was developed to aggregate the arrays resulting from each WSI into a single 64-dimensional feature array. A different exponent (denoted as $p$) of a generalized-mean function was used for each feature in a spectrum that varies between $p = 1$, when the generalized-mean is equal to the arithmetic mean, and $p = 16$, when the generalized-mean will approximate the maximum function. In mathematical terms, let $f_{ij}$ be the $j$th extracted feature of the $i$th patch, the $j$th element of output vector is $\left(\frac{1}{n}\sum_{i=1}^{n} f_{i,j}^{p_j}\right)^{1/p_j}$, where $p_j = 1 + 15(j-1)/63$. Finally, in the fourth stage, the WSI classification probability was computed using a linear layer followed by a sigmoid on the features resulting from the aggregated 64-dimensional array.

In the test phase, to decrease the patch sampling effect, the result of each WSI was evaluated 64 times, and the final probability is the mean of these values.

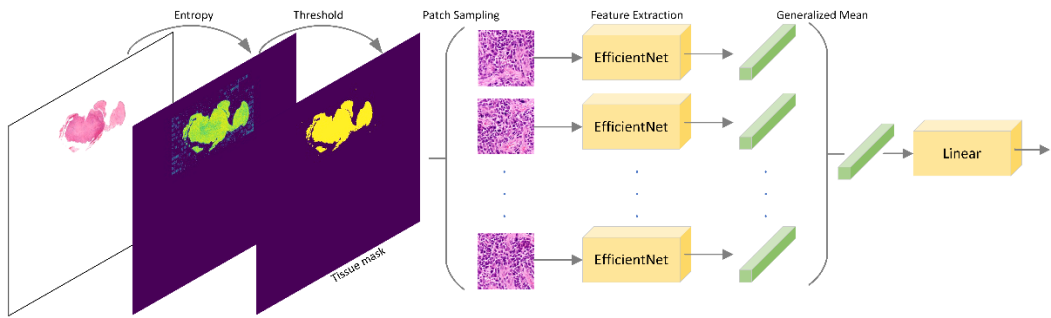

Figure 3: Overall architecture of the model developed by team Piaz

### 2.3.4. Team Dratur

The team Dratur used a method consisting of two parallel tracks (@20x and @5x tracks) to classify each WSI (see Figure 4). Tumor regions were manually annotated in 3DHistech CaseViewer and exported to TIFF file format with 3DHistech SlideConverter. After preprocessing for brightness adaption the TIFF files were sliced with a sliding window procedure generating 256 by 256 pixel tiles at the original 20x magnification and 256 by 256 pixel tiles at 5x magnification. Tiles with less than 50% of tissue pixels were discarded. A sample of tissue tiles was manually grouped for vital invasive carcinoma and non-tumor (including ductal carcinoma in situ, DCIS, in the @5x track). Two EfficientNets B4 were trained to enrich vital invasive tumors in both tracks. Strong and complex data augmentations were applied in the training of all convolutional neural networks, including the modification of hue and saturation, the addition of salt and pepper artifacts, color noise, or block artifacts using the imgaug library (https://github.com/aleju/imgaug) as well as affine augmentations from the Keras library. HER2 positive and negative cases were split into 5 partitions keeping the class balance in each partition. EfficientNets B4 and B2 were then trained using a 5-fold cross-validation procedure to predict the HER2 status. The resulting soft predictions were fed into a small dense convolutional network (two hidden layers with 32 and 16 nodes, L2 regularization, and drop out) trained with a 3-fold split for cross-validation. The resulting soft-predictions were tested against the training dataset ground-truth and a threshold of 0.47 was defined to generate the hard prediction for each WSI resulting in the correct classification of 86.39% of the training WSI, compared to a correct classification of 85.27% at a threshold of 0.5.

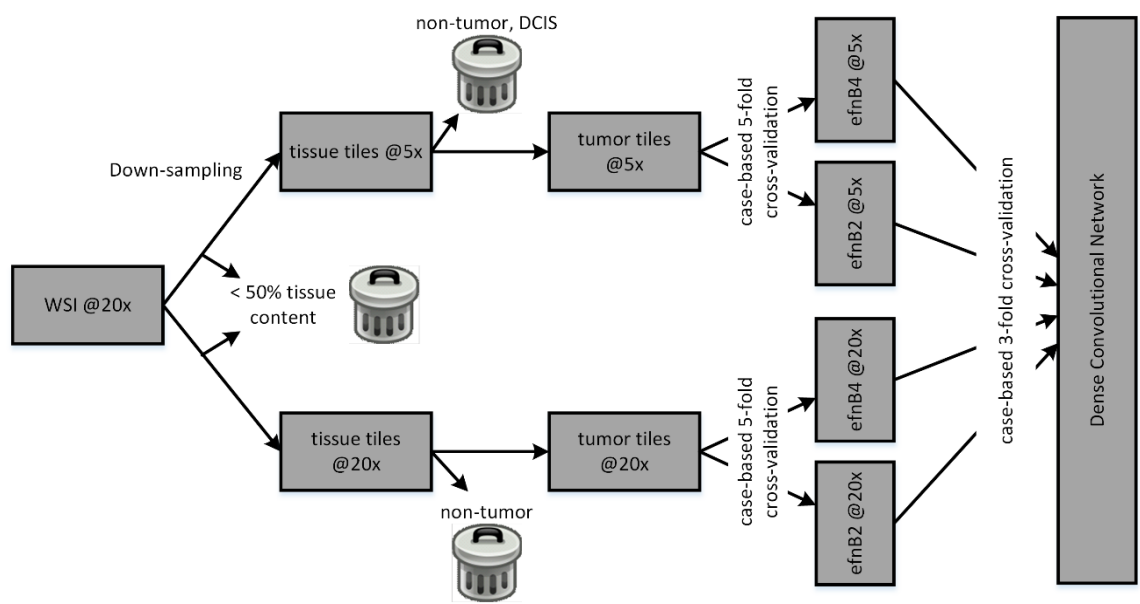

*Figure 4: Overall architecture of the model developed by team Dratur*

### 2.3.5. Team IRISAI

The team IRISAI used a two-stage model to solve the problem (see Figure 5). In the first stage, a U-Net [50] model was trained from scratch on slides at 5x magnification, to segment each WSI into "cellular" and "non-cellular" regions. Training images for this task resulted from an interactive set of annotations and corrections created in the DeePathology STUDIO software. In the second stage, 900,000 image patches of size 256x256 pixels were extracted from the cellular regions at 20x magnification and used as training dataset. A patch was considered cellular if at least 95% of its pixels were predicted as cellular by the U-Net network of stage one. Patches for the weakly supervised task were labeled by assigning to each one a value according to the corresponding WSI with a smoothing of 10% to accommodate for patch selection errors. In other words, patches originating from a positive WSI were labeled as 0.9 while those originating from a negative WSI were labeled as 0.1. Standard data augmentation was then applied to the resulting dataset and a Resnet50 classifier, pre-trained on ImageNet [47], was trained to predict the patch-level labels using an Adam optimizer with default parameters. Finally, the WSI-level HER2 score was computed by splitting it into tiles and measuring the ratio of cellular tiles in that slide that had an output of above 0.5 from the Resnet50 classifier.

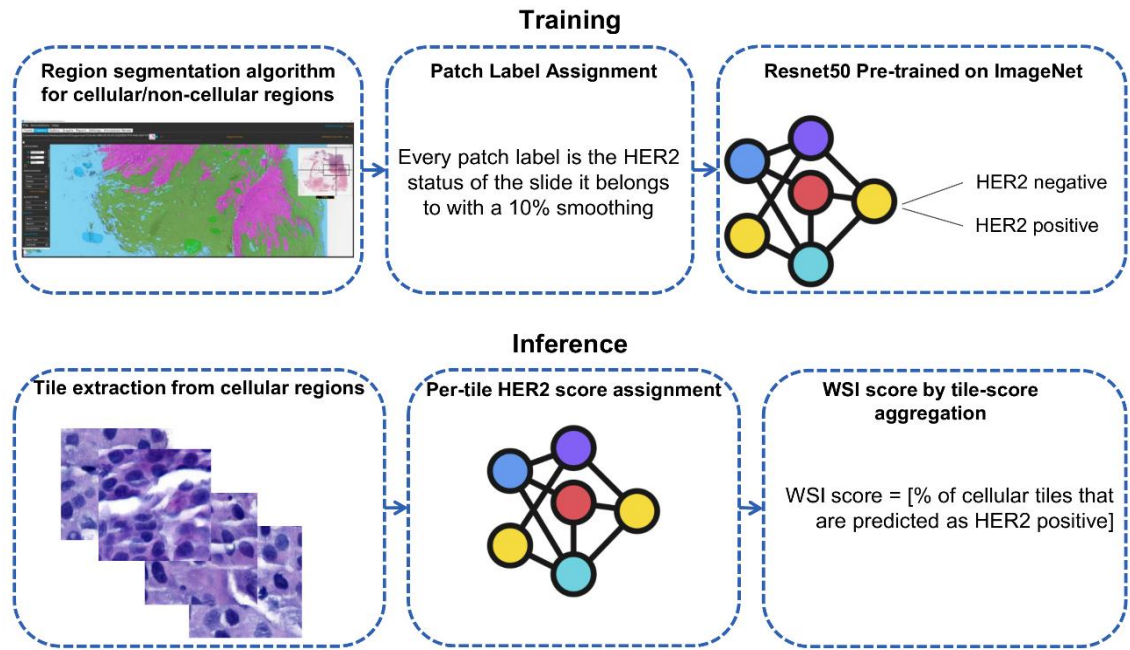

*Figure 5: Overall architecture of the model developed by team IRISAI*

### 2.3.6. Team Arontier_HYY

The team Arontier_HYY employed a three-stage method to predict HER2 status (see Figure 6). In the first stage, an EfficientNet-B3 (denoted as CNN-A) was trained from scratch to classify image patches of 1024 by 1024 pixels as "tissue" or "background". The training dataset for CNN-A consisted of image patches labeled as "background" if they include regions with stains or adipose tissue, as well as if they contain severely blurred regions and patches with less than 25% of tissue (pixel intensities bellow 240), or as "tissue" otherwise. After discarding images classified by CNN-A as "background", in the second stage, a model consisting of two parallel EfficientNets (denoted as CNN-B) was used to extract a feature vector and a score representing the probability of a given tissue patch being from a HER2 positive WSI. In one of the parallel paths, an EfficientNet-B1 was trained on image patches of 480 by 480 pixels while in the other an EfficientNet-B5 was trained on patches of size 912 by 912 pixels. Tissue patches were labeled with 0 whenever they originate from a HER2 negative WSI and with 1 otherwise. Data augmentation was then implemented to generate a more robust model and the resulting dataset was used to train CNN-B. To minimize class imbalance and inter-WSI variations, within each mini-batch, all patch images originate from different WSIs and 50% were from randomly selected positive WSI while the remaining were from randomly selected negative WSI. In stage three, all feature vectors of any given WSI were sorted in decreasing order by the corresponding score of CNN-B and fed into a Long-Short Term Memory (LSTM) network with two recurrent layers and one dropout layer to assess the final WSI prediction.

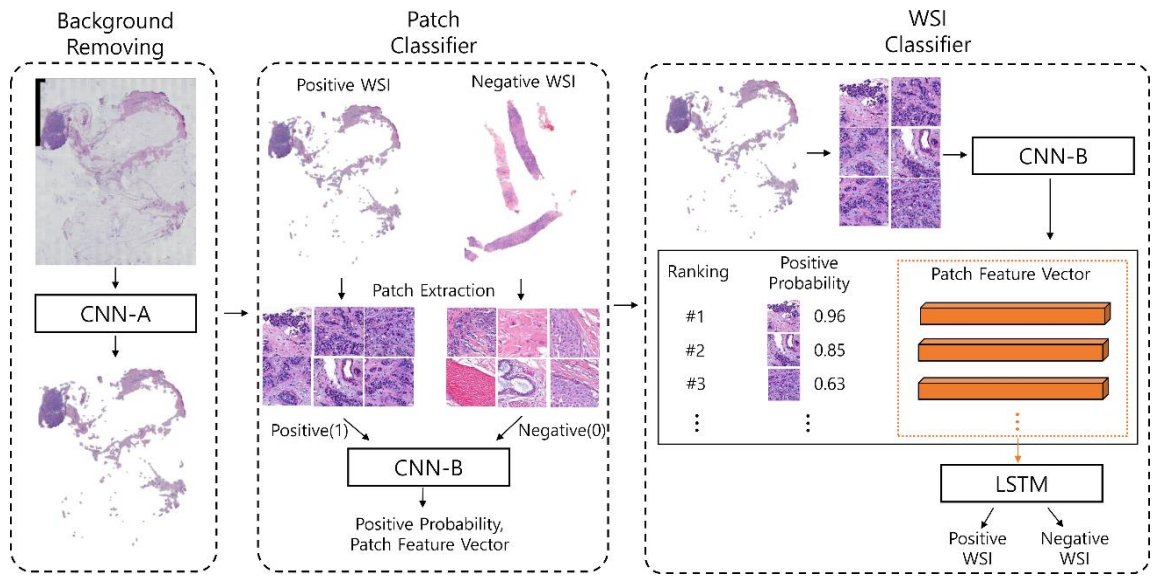

*Figure 6: Overall architecture of the model developed by team Arontier_HYY*

## 3. Results

The HEROHE Challenge was open for approximately 4 months, between October 1st, 2019, and January 28th, 2020. During this period, 863 participants registered and had access to the training and testing datasets. While many participants chose to work individually, 93 of them grouped into one of the 56 participant teams. From these, 21 teams or individual participants submitted their results and the final scores were evaluated to assign the final challenge's ranking (see Table 1).

*Table 1: Classification at the HEROHE challenge. The final classification was assessed by the $F_1$ score, but other metrics were also evaluated, namely the Area Under the Curve (AUC), Precision, and Recall.*

| RANK | TEAM | AUC | PRECISION | RECALL | $F_1$ SCORE |
|---|---|---|---|---|---|
| 1 | Macaroon | 0.71 | 0.57 | 0.83 | 0.68 |
| 2 | MITEL | 0.74 | 0.58 | 0.78 | 0.67 |
| 3 | Piaz | 0.84 | 0.77 | 0.55 | 0.64 |
| 4 | Dratur | 0.75 | 0.57 | 0.70 | 0.63 |
| 5 | IRISAI | 0.67 | 0.58 | 0.67 | 0.62 |
| 6 | Arontier_HYY | 0.72 | 0.52 | 0.73 | 0.61 |
| 7 | KDE | 0.62 | 0.51 | 0.75 | 0.61 |
| 8 | joangibert14 | 0.66 | 0.48 | 0.78 | 0.60 |
| 9 | VISILAB | 0.63 | 0.51 | 0.73 | 0.60 |
| 10 | MIRL | 0.50 | 0.40 | 1.00 | 0.57 |
| 11 | aetherAI | 0.66 | 0.49 | 0.67 | 0.57 |
| 12 | NCIC | 0.63 | 0.52 | 0.62 | 0.56 |
| 13 | biocenas | 0.57 | 0.46 | 0.53 | 0.50 |
| 14 | HEROH | 0.59 | 0.46 | 0.53 | 0.49 |
| 15 | Reza Mohebbian | 0.61 | 0.51 | 0.43 | 0.47 |
| 16 | mindmork | 0.63 | 0.53 | 0.38 | 0.45 |
| 17 | Institute of Pathology Graz | 0.63 | 0.50 | 0.38 | 0.43 |
| 18 | katherandco | 0.44 | 0.44 | 0.40 | 0.42 |
| 19 | QUILL | 0.63 | 0.50 | 0.33 | 0.40 |
| 20 | HEROHE_Challenge | 0.48 | 0.37 | 0.27 | 0.31 |
| 21 | UC-CSSE | 0.47 | 0.31 | 0.27 | 0.29 |

Two teams developed methods that, by construction, did not quantify the probability of a WSI belonging to one class (i.e., did not infer a soft prediction). The model developed by the team Macaroon classified a WSI after comparing the number of patches classified as tumor by one network and as positive by the second network. Thus, an approximation was needed to compute the AUC. The soft predictions were set to 1 whenever a WSI was classified as positive and to 0 otherwise. The model developed by team MITEL used a majority voting to assign a class to a WSI. Considering that the proposed method exported three soft predictions, the developers chose to consider as a representative soft prediction the one resulting from the overall positivity because it covers all the WSI patches, thus being considered the most exhaustive among the three.

The choice of the threshold and its impact on the $F_1$ score was assessed for the top-ranked teams by varying the threshold from 0 to 1 by steps of 0.01 and computing the best $F_1$ score on the test dataset, thus assessing the theoretical maximum of each model for the given test dataset. The results revealed that three of the top four teams could have achieved better results and the resulting ranking would have changed if different thresholds were chosen. The maximum $F_1$ scores for these teams (excluding the team Macaroon since all soft predictions were 0 or 1) were: (a) MITEL: $F_1$ = 0.70 for a threshold of 0.37; (b) Piaz: $F_1$ = 0.73 for a threshold of 0.39; and (c) Dratur: $F_1$ = 0.69 for a threshold of 0.34.

Although the $F_1$ score was the ranking metric, other metrics were also assessed. The precision and recall were assessed to compute the $F_1$ score, while the AUC was measured to allow comparisons to other recently published methods for HER2 prediction [27, 29, 30]. Two teams (MITEL and Dratur, 2[nd] and 4[th] place, respectively) showed AUC results for HER2 prediction similar to these methods while the team Piaz (3[rd] place) achieved a higher AUC of 0.84.

Considering the distribution of cases among the four possible HER2 scores on an IHC test, the precision, recall, AUC and $F_1$ score were also evaluated in the subset of equivocal IHC cases (score of 2+) and are presented in Table 2. Six teams achieved on this subset AUC equal or greater than 0.84, thus higher than the AUC achieved by the models presented in recent works [27, 29, 30].

Table 2: Results in the subset of the Test Dataset comprising just the cases classified as equivocal (score of 2+) on the IHC test.

| TEAM | AUC | PRECISION | RECALL | $F_1$ SCORE |
| --- | --- | --- | --- | --- |
| Macaroon | 0.84 | 0.75 | 0.84 | 0.79 |
| Arontier_HYY | 0.88 | 0.67 | 0.81 | 0.73 |
| MITEL | 0.85 | 0.74 | 0.72 | 0.73 |
| Dratur | 0.85 | 0.71 | 0.75 | 0.73 |
| IRISAI | 0.85 | 0.72 | 0.72 | 0.72 |
| KDE | 0.77 | 0.67 | 0.75 | 0.71 |
| Piaz | 0.84 | 0.79 | 0.59 | 0.68 |
| VISILAB | 0.77 | 0.64 | 0.66 | 0.65 |
| NCIC | 0.70 | 0.58 | 0.69 | 0.63 |
| biocenas | 0.71 | 0.61 | 0.63 | 0.62 |
| aetherAI | 0.77 | 0.53 | 0.72 | 0.61 |
| QUILL | 0.78 | 0.79 | 0.47 | 0.59 |
| joangibert14 | 0.70 | 0.46 | 0.72 | 0.56 |
| MIRL | 0.50 | 0.38 | 1.00 | 0.55 |
| Reza Mohebbian | 0.64 | 0.52 | 0.47 | 0.49 |
| Institute of Pathology Graz | 0.70 | 0.50 | 0.47 | 0.48 |
| HEROH | 0.63 | 0.46 | 0.50 | 0.48 |
| mindmork | 0.61 | 0.43 | 0.31 | 0.36 |
| katherandco | 0.32 | 0.67 | 0.25 | 0.36 |

| | | | | | |
|---|---|---|---|---|---|
| UC-CSSE | | 0.61 | 0.42 | 0.31 | 0.36 |
| HEROHE_Challenge | | 0.50 | 0.37 | 0.22 | 0.27 |

Although the core methodologies differ between submitted models, some procedures are common among approaches. For example, 20 out of the 21 teams developed methods taking advantage of deep neural networks in one or more steps of their models and used Python as the main programming language. Reza Mohebbian was the only team that developed a "classical" Machine Learning model (without deep neural networks) coded on MATLAB. Teams HEROH, HEROHE_Challenge, Institute of Pathology Graz, and katherandco, despite having used Python as the main programing language, used also QuPath [51] on some steps of their methods. Another common step was the split of each WSI into smaller patches. Only three teams used the entire WSI as input, Reza Mohebbian, that developed a non-deep learning-based model; Institute of Pathology Graz, that combined hand-crafted feature extractor developed on QuPath with a custom CNN for classification; and aetherAI, that adapted a ResNet50 [28] to receive as input WSIs resized to a 10.000 by 10.000 pixels canvas (see 2.3. Competing Solutions, and Supplementary Material for more details). Among, the 20 teams that used deep neural networks on their methods, 12 chose to rely on models pre-trained on other publicly available datasets (more details in Table 3), being the ImageNet [47] the most widespread, but other datasets were used [34, 35, 52, 53].

Table 3: Summary of submitted methods. **Approach** lists the main methods to classify the WSI; **Pre-trained** indicates if the transfer learning approach was used; **Ensemble** indicates if the method uses one or multiple models and their number; **External sets** indicates external datasets used in pre-trained models; **Input size** indicates the size, in pixels, of the images or tiles required by the model (WSI means that the entire WSI is passed to the model at the same time).

| RANK | TEAM | APPROACH | PRE-TRAINED | ENSEMBLE | EXTERNAL SETS | INPUT SIZE |
|---|---|---|---|---|---|---|
| 1 | Macaroon | ResNet34 | yes | 2 | CAMELYON16 | 256x256 |
| 2 | MITEL | DenseNet201 + ResNet152 | yes | 2 | ImageNet + BACH | 512x512 |
| 3 | Piaz | EfficientNet-B0 | yes | x | BACH | 222x222 |
| 4 | Dratur | EfficientNetB2 + EfficientNetB4 + Custom dense model | yes | 5 | ImageNet | 256x256 |
| 5 | IRISAI | U-Net + ResNet50 | no + yes | 2 | ImageNet | 256x256 |
| 6 | Arontier_HYY | EfficientNetB1 + EfficientNetB3 + EfficientNetB5 + LSTM | no | 4 | x | 1024x1024 + 480x840 + 912x912 |
| 7 | KDE | Custom + InceptionV3 | no | 3 | x | 128x128 |
| 8 | joangibert14 | ResNet101 | yes | x | [52] | 224x224 |
| 9 | VISILAB | SE-ResNet50 | no | x | | 299x299 |
| 10 | MIRL | DenseNet201 | yes | x | ImageNet | 9192 x 9192 |
| 11 | aetherAI | Custom based on ResNet 50 v2 | no | x | x | WSI rescaled to 10000x10000 |
| 12 | NCIC | Resnext101 + Resnext50 [54] | yes | 2 | ImageNet | 1024x1024 |
| 13 | biocenas | Custom CNN model | no | 3 | x | 32x32 |
| 14 | HEROH | ResNet18 + ResNet50 | yes | 2 | ImageNet | 128x128 |
| 15 | Reza Mohebbian | Custom (non-Deep Learning) | no | x | x | WSI |
| 16 | mindmork | Kmeans [55, 56] + U-Net+Xception [57] | no | 3 | x | 256x256 |
| 17 | Institute of Pathology Graz | QuPath for color deconvolution and feature extractor + Custom CNN | no | 2 | x | WSI |
| 18 | katherandco | QuPath for tumor segmentation + ResNet50 | no | x | ImageNet | 512x512 |

| 19 | QUILL | SuperPixel patch spliting + DenseNet + Mean Shift Clustering | no | 2 | x | WSI |
| 20 | HEROHE_Challenge | Custom CNN + kmeans + XGBoost | yes | 3 | CIFAR-10 dataset | 200x200 |
| 21 | UC-CSSE | Xception + densenet169 + resnet34 + resnet101 + random forest + extra trees + gradient boosting | yes | 7 | CAMELYON16 + Data Science Bowl 2018 | 299x299 |

## 4. Discussion

As in all challenges, the definition of the metric to access the final ranking is of paramount importance. All metrics that may have been considered to evaluate the methods assess different aspects of the results so they would produce different ranks. Since our dataset was imbalanced, metrics such as accuracy (the percentage of total cases correctly classified) that is often used to evaluate classifiers is not a reliable descriptor of the model's adequacy to solve the problem at hand. For example, a model that predicts all cases as negative achieves an accuracy equal to the proportion of the negative cases, which, for highly imbalanced datasets, would result in higher classifications. Other metrics are less prone to the class imbalance problem. Among them, precision and recall are good choices nevertheless both these metrics fail in some extreme scenarios. For example, the team MIRL, ranked tenth on the challenge (Table 1), achieved a perfect recall. Looking in more detail for the team MIRL's predictions (Table 4) one fact can be highlighted, team MIRL predicted all cases as positives. This, while guaranteeing that no false negatives would result from the predictions, and thus the recall would be 1, leads to a model that, despite having one perfect metric, is useless to solve the problem here proposed. Cases as this one can be pinpointed by comparing the model's recall with its precision. Team MIRL achieved a precision of 0.4, a value that corresponds to the proportion of positive cases on the test dataset (Table 1). Analogously, if a model only predicts as positive a very small percentage of the true positives, but none of the negatives, it will have a precision of 1, while failing to properly predict most of the relevant cases, thus having a small recall. The proper balance between these two metrics can be achieved by combining them into a single and more robust metric, the $F_1$ score, which only achieves values close to 1 if both the precision and the recall are simultaneously close to 1 (e.g. team MIRL had a recall of 1 but a precision of 0.4, which resulted into a $F_1$ score of 0.57).

Table 4: Confusion matrix of team MIRL. $tp$: true positives; $tn$: true negatives; $fp$: false positives; $fn$: false negatives.

|  |  | Predicted Condition | | Total |
|---|---|---|---|---|
|  |  | Positive | Negative |  |
| **Actual Condition** | Positive | $tp = 60$ | $fn = 0$ | 60 |
|  | Negative | $fp = 90$ | $tn = 0$ | 90 |
|  | Total | 150 | 0 | 150 |

The organizers did not opt for other performance metrics, such as the AUC, to better simulate the clinical practice. Indeed, AUC is a global performance metric that does not necessarily encode the behavior of a system at different regions of the ROC curve. For instance, in clinical practice it may be of interest to operate on a region of lower or higher FPR, depending on the goal of the screening. However, two models with the same AUC may exhibit different behaviors in these two extreme regions. With these in mind, it was decided to instead ask participants to select a prediction threshold that would maximize the $F_1$ score. The main goal was to force the system to make an absolute/objective decision regarding the sample being assessed, serving as a clear second opinion about the patient's HER2 status, i.e. the participants were asked *a priori* to select the operation point of their system.

Previous studies addressing the problem of predicting the expression of molecular biomarkers in breast cancer [27, 29, 30] reported as primary performance metrics the AUC. Given that, despite having considered the $F_1$ score as the ranking metric for the challenge, the AUC was used here as the metric for performance assessment. The models presented in section 2.3. Competing Solutions correspond to those that achieved AUC > 0.8 in Table 1 or in Table 2, i.e. AUC bigger than those reported on the referred studies. The remaining methods are described in the Supplementary Material and a summary of all models is presented in Table 3.

An important aspect that needs to be considered by all teams willing to participate in challenges is the decision on the threshold value to decide the final prediction. If the teams had chosen different thresholds to generate their final predictions, their results in the $F_1$ score could have been better. For instance, with this test dataset, the model developed by the team Piaz could have achieved a $F_1$ score of 0.73 (instead of the actual 0.64) if the threshold was set to 0.39 (instead of the used 0.5), thus being ranked at the top of the leaderboard. Of course, the value 0.39 was obtained as the best choice for the test dataset, which, by definition, was not available to teams. Nevertheless, the training dataset or part of it (e.g., validation dataset) should be used to fine-tune the threshold. It is worth noting that, even if the optimum threshold was not the one resulting by the evaluation of the model performance on the training dataset, others close to 0.39 (thresholds between 0.36 and 0.43) would also generate a final $F_1$ score greater than 0.70.

The comparison between the result that team MITEL reached ($F_1 = 0.67$) and the one that they would have had if they used the overall connectivity as the only evaluation feature ($F_1 = 0.70$ for a threshold of 0.37) reveals that sometimes a more complex model does not outperform a simpler one.

The results obtained by the teams in the subset of equivocal cases (Table 2) revealed that most of the top-ranked teams (8 of the top 10, including the top 7 teams) achieved better results on this subset than in the whole test dataset (Table 1). For example, team Macaroon achieved a $F_1$ score of 0.68 on the whole test dataset, but this figure raised to 0.79 when considering just the equivocal cases. This difference is due to the increase in precision that raises from 0.57 to 0.75 while the recall increases from 0.83 to 0.84. Similar changes occurred with other top-ranked teams (e.g. Dratur's $F_1$ score raised from 0.63 to 0.73, Arontier_HYY's $F_1$ score raised from 0.61 to 0.73). Equivalently, the AUC increased in most of the teams. While on the full test dataset only one team achieved an AUC greater than 0.8 (team Piaz with $AUC = 0.84$), on the subset of the equivocal cases 6 teams had AUC above 0.8. The difference between the results achieved on these two groups of cases can be explained by the distribution of cases per HER2 score on both the training and test datasets with a majority being classified as equivocal on the IHC test.

Following a common trend in medical image analysis, the vast majority of the teams (20 out of 21) used at least one deep neural network on their models. Given the complexity of the task, many teams split the problem into more than one step and rely on the combination of more than one deep network to classify each WSI (Table 3). Since the training dataset was comprised of 359 cases, 12 teams chose to use models pre-trained on other datasets. This allowed them to train deeper networks, with up to millions of parameters, and potentially improve performance while reducing development and training time. Another common technique was splitting the entire WSI into small tiles (18 out of 21 teams), applying the HER2 classifier to each tile, and later combining the information to get the WSI classification. Most teams (16 out of 21) decided to prune the WSI first by applying a segmentation algorithm that identifies regions of interest (e.g. the two top-ranked teams applied DL algorithms to identify tumor regions while the third-ranked team relied on non-DL algorithms to segment tissue regions) and only then use those regions in the classification step, thus reducing the computational costs, while focusing the WSI classification on targeted regions.

In terms of clinical application, ideally, the step to take further comprehends not only the prediction of the HER2 status in BC samples but also predicting the response of the patients to HER2 targeted therapy. Previous literature shows that morphological clues can be found in the tumor, such as the presence of tumor-infiltrating lymphocytes (TILs), which can be good predictors of HER2 target therapy. BC samples with a high number of TILs will display more often complete pathological responses in the surgical specimen after neoadjuvant (before surgery) HER2 target therapy and subsequent better disease-free survival [58, 59]. Moreover, there might be additional features that could be extracted for predicting the response to HER2 target therapy.

In specific settings of breast cancer, gene expression tests have already been recommended to assess the risk of recurrence and guide oncologists in the difficult decision to use chemotherapy [60, 61]. These tests are very expensive and tissue destructive, two major limitations that decrease their use in clinical practice. Surprisingly or not, it has been shown that nuclear morphology features, such as nuclear shape and architecture, can be extracted from H&E stained images to predict risk categories using the gene signatures [62]. This type of research and prediction can take computational pathology to a level never experienced in medicine.

In the HEROHE Challenge, it was decided to consider datasets with cases without HER2 heterogeneity since it is the most frequent situation. Additionally, heterogeneous cases would require specific tumor annotations. HER2 heterogeneity corresponds to tumors with both HER2-negative and HER2-positive areas, representing a minority of situations (up to 1% of the cases) with patients requiring at least 10% of HER2-positive areas in the BC to be elected for targeted therapy [4, 63-65]. Although there are morphological features more likely to be associated with HER2-positive BC (as discussed above), making the distinction between HER2-positive and HER2-negative theoretically possible, in cases with HER2 heterogeneity the different areas in the BC appear to be very similar, at least to the pathologist´s assessment [64]. The proportion of cases in each IHC score was biased towards the class 2+. This decision was because these are the cases that require further assessment, namely by the evaluation of HER2 amplification with ISH, and thus considered by the challenge organizers as the most important cases. Future research to address this problem should consider these aspects.

Roughly, 15% of the cases evaluated by IPATIMUP diagnostics are positives, nevertheless, the datasets here released failed to follow the real ratio between positive and negative cases. This decision was taken based on the logistics required to release a dataset according to this ratio that still had a sufficient number of WSI per class. Such a dataset would require more than 3TB of disk space and would result in a dramatic increase in the communication time required to upload and download all the data. On one hand, such a huge dataset could eventually prevent some teams from participating due to the computational resources that would become necessary. On the other hand, more data may have resulted in better models, although the higher class imbalance could also present different challenges to teams during model development and training.

## 5. Conclusions

The HEROHE Challenge was developed with the primary goal of promoting the development of computer-aided diagnosis tools to predict the HER2 status in invasive BC samples. Despite the complexity of the proposed task, 21 models were presented, combining different techniques, from standard image analysis to state-of-the-art DL algorithms, and promising results were achieved. Two of the top-ranked teams achieved AUC similar to those presented in other recent works while one team achieved a higher AUC.

Given the biased distribution of training and test datasets, with a majority of cases classified as equivocal by the IHC test, this work also presented the AUC and $F_1$ score of the proposed models in a subset of the test dataset with only equivocal cases (Table 2). Most of the teams performed better on this subsample. Six teams achieved AUC greater than or equal to 0.84, clearly outperforming the results presented in recently published studies. Team Arontier_HYY achieved the higher AUC (0.88) on this dataset and the top $F_1$ score raised from 0.69 to 0.79 (team Macaroon). This fact suggests that some of the presented models identified features on the H&E slides scored of 2+ on the IHC test that can be used to predict HER2 status without access to IHC or ISH assays, something that human experts are not able to do. The achieved results were not perfect and more data may lead to an improvement in the performance of the models, especially in cases scored as 0, 1+ or 3+ by the IHC test that were under-represented in the challenge datasets, but eventually in the equivocal cases also.

The importance of the metric defined to assess the models' performance was shown to have a great impact on the final rank. In this work, the importance of a proper selection of the final threshold to separate positive and negative cases was also emphasized by presenting an example revealing that with a better choice of the threshold, the same algorithm would result in a model with a significantly better performance. It was also pointed that the choice of the network, hyperparameters, and all the features of a model has to be carefully evaluated during the development, but also later, taking into consideration the evaluation of the model on the training and validation datasets. The result of the team MITEL compared to the one they could have had with a simpler model is a clear remark that is important to keep in mind.

Although the challenge is now closed, the website and datasets will remain public for research purposes, thus further contributing to the development of novel solutions to automate HER2 status assessment.

## Funding

Eduardo Conde-Sousa is supported by a post-doctoral grant of the project PPBI-POCI-01-0145-FEDER-022122, in the scope of FCT National Roadmap of Research Infrastructures.

Guilherme Aresta is funded by the FCT grant contract SFRH/BD/120435/2016.

Teresa Araújo is funded by the FCT grant contract SFRH/BD/122365/2016.

## Declaration of Competing Interest

Jacob Gildenblat (JG) is the Co-Founder and CTO of DeePathology that collaborates with Roche on deep learning for digital pathology. Eldad Klaiman (EK) is affiliated to Roche Diagnostics. Despite ROCHE sponsored the challenge, the participation of JG and EK as members of the team IRISAI followed the same rules as all the other participants.

The authors declare that they have no known competing financial interests or personal relationships that could have appeared to influence the work reported in this paper.

## Acknowledgments

The HEROHE Challenge was an initiative of the European Society on Digital and Integrative Pathology (ESDPI) during the organization of the European Congress on Digital Pathology 2020.

The authors thank ROCHE for supporting the HEROHE Challenge.

The authors thank all the HEROHE participants that submitted their methods.

# References

1. Bray, F., et al., *Global cancer statistics 2018: GLOBOCAN estimates of incidence and mortality worldwide for 36 cancers in 185 countries.* CA Cancer J Clin, 2018. **68**(6): p. 394-424.
2. Creyten, D. and U. Flucke, *WHO classification of tumours : breast tumours*. Vol. 2. 2020: International Agency for Research on Cancer (IARC).
3. Allison, K.H., et al., *Estrogen and Progesterone Receptor Testing in Breast Cancer: American Society of Clinical Oncology/College of American Pathologists Guideline Update.* Arch Pathol Lab Med, 2020. **144**(5): p. 545-563.
4. Wolff, A.C., et al., *Human Epidermal Growth Factor Receptor 2 Testing in Breast Cancer: American Society of Clinical Oncology/College of American Pathologists Clinical Practice Guideline Focused Update.* Arch Pathol Lab Med, 2018. **142**(11): p. 1364-1382.
5. Slamon, D.J., et al., *Human breast cancer: correlation of relapse and survival with amplification of the HER-2/neu oncogene.* Science, 1987. **235**(4785): p. 177-82.
6. Press, M.F., et al., *HER-2/neu gene amplification characterized by fluorescence in situ hybridization: poor prognosis in node-negative breast carcinomas.* J Clin Oncol, 1997. **15**(8): p. 2894-904.
7. Slamon, D.J., et al., *Use of chemotherapy plus a monoclonal antibody against HER2 for metastatic breast cancer that overexpresses HER2.* N Engl J Med, 2001. **344**(11): p. 783-92.
8. Vogel, C.L., et al., *Efficacy and safety of trastuzumab as a single agent in first-line treatment of HER2-overexpressing metastatic breast cancer.* J Clin Oncol, 2002. **20**(3): p. 719-26.
9. Piccart-Gebhart, M.J., et al., *Trastuzumab after adjuvant chemotherapy in HER2-positive breast cancer.* N Engl J Med, 2005. **353**(16): p. 1659-72.
10. Gruver, A.M., Z. Peerwani, and R.R. Tubbs, *Out of the darkness and into the light: bright field in situ hybridisation for delineation of ERBB2 (HER2) status in breast carcinoma.* J Clin Pathol, 2010. **63**(3): p. 210-9.
11. Hariri, N., et al., *Cost-effectiveness of a Dual (Immunohistochemistry and Fluorescence In Situ Hybridization) HER2/neu Testing Strategy on Invasive Breast Cancers.* Appl Immunohistochem Mol Morphol, 2020.
12. Mukhopadhyay, S., et al., *Whole Slide Imaging Versus Microscopy for Primary Diagnosis in Surgical Pathology: A Multicenter Blinded Randomized Noninferiority Study of 1992 Cases (Pivotal Study).* Am J Surg Pathol, 2018. **42**(1): p. 39-52.
13. Stathonikos, N., et al., *Going fully digital: Perspective of a Dutch academic pathology lab.* J Pathol Inform, 2013. **4**: p. 15.
14. Thorstenson, S., J. Molin, and C. Lundstrom, *Implementation of large-scale routine diagnostics using whole slide imaging in Sweden: Digital pathology experiences 2006-2013.* J Pathol Inform, 2014. **5**(1): p. 14.
15. Snead, D.R., et al., *Validation of digital pathology imaging for primary histopathological diagnosis.* Histopathology, 2016. **68**(7): p. 1063-72.
16. Cheng, C.L., et al., *Enabling digital pathology in the diagnostic setting: navigating through the implementation journey in an academic medical centre.* J Clin Pathol, 2016. **69**(9): p. 784-92.
17. Araujo, A.L.D., et al., *The performance of digital microscopy for primary diagnosis in human pathology: a systematic review.* Virchows Arch, 2019. **474**(3): p. 269-287.
18. Hartman, D.J., et al., *Enterprise Implementation of Digital Pathology: Feasibility, Challenges, and Opportunities.* J Digit Imaging, 2017. **30**(5): p. 555-560.
19. Kowal, M., et al., *Computer-aided diagnosis of breast cancer based on fine needle biopsy microscopic images.* Comput Biol Med, 2013. **43**(10): p. 1563-72.
20. George, Y.M., et al., *Remote Computer-Aided Breast Cancer Detection and Diagnosis System Based on Cytological Images.* IEEE Systems Journal, 2014. **8**(3): p. 949-964.
21. Zhang, B. *Breast cancer diagnosis from biopsy images by serial fusion of Random Subspace ensembles*. in *2011 4th International Conference on Biomedical Engineering and Informatics (BMEI)*. 2011.
22. Fondon, I., et al., *Automatic classification of tissue malignancy for breast carcinoma diagnosis.* Comput Biol Med, 2018. **96**: p. 41-51.
23. Araújo, T., et al., *Classification of breast cancer histology images using Convolutional Neural Networks.* PLOS ONE, 2017. **12**(6): p. e0177544.
24. Vandenberghe, M.E., et al., *Relevance of deep learning to facilitate the diagnosis of HER2 status in breast cancer.* Sci Rep, 2017. **7**: p. 45938.
25. Khameneh, F.D., S. Razavi, and M. Kamasak, *Automated segmentation of cell membranes to evaluate HER2 status in whole slide images using a modified deep learning network.* Computers in Biology and Medicine, 2019. **110**: p. 164-174.
26. Couture, H.D., et al., *Image analysis with deep learning to predict breast cancer grade, ER status, histologic subtype, and intrinsic subtype.* NPJ Breast Cancer, 2018. **4**: p. 30.
27. Shamai, G., et al., *Artificial Intelligence Algorithms to Assess Hormonal Status From Tissue Microarrays in Patients With Breast Cancer.* JAMA Netw Open, 2019. **2**(7): p. e197700.
28. He, K., et al. *Deep Residual Learning for Image Recognition*. in *2016 IEEE Conference on Computer Vision and Pattern Recognition (CVPR)*. 2016.
29. Naik, N., et al., *Deep learning-enabled breast cancer hormonal receptor status determination from base-level H&E stains.* Nat Commun, 2020. **11**(1): p. 5727.

30. Kather, J.N., et al., *Pan-cancer image-based detection of clinically actionable genetic alterations.* Nature Cancer, 2020. **1**(8): p. 789-799.
31. Zhang, X., et al. *Shufflenet: An extremely efficient convolutional neural network for mobile devices*. in *Proceedings of the IEEE conference on computer vision and pattern recognition*. 2018.
32. Qaiser, T., et al., *HER2 challenge contest: a detailed assessment of automated HER2 scoring algorithms in whole slide images of breast cancer tissues.* Histopathology, 2018. **72**(2): p. 227-238.
33. Litjens, G., et al., *1399 H&E-stained sentinel lymph node sections of breast cancer patients: the CAMELYON dataset.* Gigascience, 2018. **7**(6).
34. Ehteshami Bejnordi, B., et al., *Diagnostic Assessment of Deep Learning Algorithms for Detection of Lymph Node Metastases in Women With Breast Cancer.* JAMA, 2017. **318**(22): p. 2199-2210.
35. Aresta, G., et al., *BACH: Grand challenge on breast cancer histology images.* Med Image Anal, 2019. **56**: p. 122-139.
36. Hartman, D.J., et al., *Value of Public Challenges for the Development of Pathology Deep Learning Algorithms.* J Pathol Inform, 2020. **11**: p. 7.
37. Abd El-Rehim, D.M., et al., *High-throughput protein expression analysis using tissue microarray technology of a large well-characterised series identifies biologically distinct classes of breast cancer confirming recent cDNA expression analyses.* Int J Cancer, 2005. **116**(3): p. 340-50.
38. Carey, L.A., et al., *Race, breast cancer subtypes, and survival in the Carolina Breast Cancer Study.* JAMA, 2006. **295**(21): p. 2492-502.
39. Tamimi, R.M., et al., *Comparison of molecular phenotypes of ductal carcinoma in situ and invasive breast cancer.* Breast Cancer Res, 2008. **10**(4): p. R67.
40. Jackson, C.R., A. Sriharan, and L.J. Vaickus, *A machine learning algorithm for simulating immunohistochemistry: development of SOX10 virtual IHC and evaluation on primarily melanocytic neoplasms.* Mod Pathol, 2020. **33**(9): p. 1638-1648.
41. Lahiani, A., et al., *Seamless Virtual Whole Slide Image Synthesis and Validation Using Perceptual Embedding Consistency.* IEEE J Biomed Health Inform, 2021. **25**(2): p. 403-411.
42. Xu, Z., et al., *GAN-based virtual re-staining: a promising solution for whole slide image analysis.* arXiv preprint arXiv:1901.04059, 2019.
43. Fawcett, T., *An introduction to ROC analysis.* Pattern Recognition Letters, 2006. **27**(8): p. 861-874.
44. Tian, Y., et al., *Computer-aided Detection of Squamous Carcinoma of the Cervix in Whole Slide Images.* ArXiv, 2019. **abs/1905.10959**.
45. La Barbera, D., et al., *Detection of HER2 from Haematoxylin-Eosin Slides Through a Cascade of Deep Learning Classifiers via Multi-Instance Learning.* Journal of Imaging, 2020. **6**(9).
46. Huang, G., et al. *Densely connected convolutional networks*. in *Proceedings of the IEEE conference on computer vision and pattern recognition*. 2017.
47. Deng, J., et al. *ImageNet: A large-scale hierarchical image database*. in *2009 IEEE Conference on Computer Vision and Pattern Recognition*. 2009.
48. Glasbey, C.A., *An Analysis of Histogram-Based Thresholding Algorithms.* CVGIP: Graphical Models and Image Processing, 1993. **55**(6): p. 532-537.
49. Tan, M. and Q. Le, *EfficientNet: Rethinking Model Scaling for Convolutional Neural Networks*, in *Proceedings of the 36th International Conference on Machine Learning*, C. Kamalika and S. Ruslan, Editors. 2019, PMLR: Proceedings of Machine Learning Research. p. 6105--6114.
50. Ronneberger, O., P. Fischer, and T. Brox, *U-Net: Convolutional Networks for Biomedical Image Segmentation*. 2015. Cham: Springer International Publishing.
51. Bankhead, P., et al., *QuPath: Open source software for digital pathology image analysis.* Sci Rep, 2017. **7**(1): p. 16878.
52. Campanella, G., et al., *Clinical-grade computational pathology using weakly supervised deep learning on whole slide images.* Nat Med, 2019. **25**(8): p. 1301-1309.
53. Krizhevsky, A., *Learning Multiple Layers of Features from Tiny Images.* University of Toronto, 2012.
54. Xie, S., et al. *Aggregated residual transformations for deep neural networks*. in *Proceedings of the IEEE conference on computer vision and pattern recognition*. 2017.
55. Lloyd, S., *Least square quantization in PCM. Bell Telephone Laboratories Paper. Published in journal much later: Lloyd, SP: Least squares quantization in PCM.* IEEE Trans. Inform. Theor.(1957/1982), 1957. **18**.
56. MacQueen, J. *Some methods for classification and analysis of multivariate observations*. in *Proceedings of the fifth Berkeley symposium on mathematical statistics and probability*. 1967. Oakland, CA, USA.
57. Chollet, F. *Xception: Deep learning with depthwise separable convolutions*. in *Proceedings of the IEEE conference on computer vision and pattern recognition*. 2017.
58. Lee, H.J., et al., *Prognostic Significance of Tumor-Infiltrating Lymphocytes and the Tertiary Lymphoid Structures in HER2-Positive Breast Cancer Treated With Adjuvant Trastuzumab.* Am J Clin Pathol, 2015. **144**(2): p. 278-88.
59. Salgado, R., et al., *Tumor-Infiltrating Lymphocytes and Associations With Pathological Complete Response and Event-Free Survival in HER2-Positive Early-Stage Breast Cancer Treated With Lapatinib and Trastuzumab: A Secondary Analysis of the NeoALTTO Trial.* JAMA Oncol, 2015. **1**(4): p. 448-54.

60. Harris, L.N., et al., *Use of Biomarkers to Guide Decisions on Adjuvant Systemic Therapy for Women With Early-Stage Invasive Breast Cancer: American Society of Clinical Oncology Clinical Practice Guideline.* J Clin Oncol, 2016. **34**(10): p. 1134-50.
61. Krop, I., et al., *Use of Biomarkers to Guide Decisions on Adjuvant Systemic Therapy for Women With Early-Stage Invasive Breast Cancer: American Society of Clinical Oncology Clinical Practice Guideline Focused Update.* J Clin Oncol, 2017. **35**(24): p. 2838-2847.
62. Whitney, J., et al., *Quantitative nuclear histomorphometry predicts oncotype DX risk categories for early stage ER+ breast cancer.* BMC Cancer, 2018. **18**(1): p. 610.
63. Polonia, A., D. Leitao, and F. Schmitt, *Application of the 2013 ASCO/CAP guideline and the SISH technique for HER2 testing of breast cancer selects more patients for anti-HER2 treatment.* Virchows Arch, 2016. **468**(4): p. 417-23.
64. Polonia, A., G. Oliveira, and F. Schmitt, *Characterization of HER2 gene amplification heterogeneity in invasive and in situ breast cancer using bright-field in situ hybridization.* Virchows Arch, 2017. **471**(5): p. 589-598.
65. Curado, M., et al., *What to expect from the 2018 ASCO/CAP HER2 guideline in the reflex in situ hybridization test of immunohistochemically equivocal 2+ cases?* Virchows Arch, 2019. **475**(3): p. 303-311.

# Supplementary Material of "HEROHE Challenge: assessing HER2 status in breast cancer without immunohistochemistry or in situ hybridization"

## 7 – KDE

The team KDE used a model that combined three deep learning modules. The first network aims to identify tumor regions, the second to extract features, and the last to classify the WSI. To segment tumor regions, a low-resolution version of the WSI was considered and an Otsu threshold and median filter were applied to identify the tissue regions. A custom convolutional neural network was trained on a manually annotated dataset to classify a patch as tumor or non-tumor, and tissue regions were classified by this network to identify tumor patches. Non-tumor patches are discarded in this stage. The remaining patches were then randomly split into five groups and a 5-fold cross-validation setup was implemented using an InceptionV3 network [1] as a feature extractor. For the third stage, an end-to-end trainable multiple instance learning architecture, called KDE [2], was trained with a 10-fold cross-validation setup on the feature vectors resulting from stage 2. The final WSI score results from the average score in each of the 10 KDE models.

## 8 – joangibert14

The team joangibert14 developed a method based on Campanella and colleagues' work [3]. After tiling the WSI into 224x224 pixel patches and removing patches corresponding to non-tissue regions, data augmentations consisting of random horizontal and vertical flips, rotations, changes in brightness and contrast, and greyscale levels were applied to enlarge the training dataset. A ResNet101 [4] model was then trained to classify a WSI based on the Multiple Instance Learning (MIL) [5] algorithm described in [3].

## 9 – VISILAB

The team VISILAB developed a model based on the SE-ResNet50 architecture [6]. The method starts by preprocessing the input images. Each WSI is split into 299 by 299 pixel patches that are then segmented to separate tissue from background patches. Image patches are considered as tissue, and used for the next steps, if and only if the mean intensity is below $0.96 \times 255 = 244.8$. Background patches are discarded. Data augmentation, including random rotations, cropping and resizing, are then applied to the resulting dataset before training the network.

## 10 – MIRL

The team MIRL developed a method consisting of pre-processing and application of a convolutional neural network for classification. The pre-processing step consisted in split each WSI into patches of 9192 x 9192 pixels and application of a median filter followed by thresholding to identify tissue regions. If the patch contained more than 30% of tissue pixels then it was tiled into 224 by 224 pixel patches and 100 random patches kept per WSI. Then, a DenseNet201 [7] pre-trained on the ImageNet [8] was retrained to predict HER2 status in each patch and a Majority Voting rule was applied to obtain the final WSI classification.

## 11 – aetherAI

The team aetherAI adapted a ResNet50 [4] model so it can support 10.000 by 10.000 pixels images as input. To be able to train a model with such memory requirements, a computational pipeline leveraging the NVIDIA Unified Memory technology was developed enabling GPU to directly access system memory, thus avoiding out-of-memory errors during training. In addition, batch normalization layers in the ResNet50 model were replaced by group normalization layers since the latter allowed for the implementation of the distributed computing system. Data pre-processing consisted of two steps, first WSIs were cropped to remove non-tissue regions, and then resized to a 10.000x10.000 pixels canvas. Data augmentation by application of random rotations and translations was also implemented to enlarge the training dataset.

## 12 – NCIC

The team NCIC created a model based on two parallel networks, a resnext101 and resnext50 [9], combined to access the WSI classification. The method starts by pre-processing each WSI. A downsampled version of it (ratio 1/16) was considered and an Otsu threshold was applied to segment tissue regions. Tissue regions were then split into 1024 x 1024 pixel patches at full resolution. Each of the parallel networks was trained according to the following protocol:

1. First, patches from a given WSI were combined to form a bag.
2. For each iteration of the training protocol, an inference is performed and $K = 512$ patches are selected from each bag and labeled according to the corresponding slide label for supervised back-propagation training. Three strategies were considered to select patches:
    i. Max-Max: select the higher scored $K$ patches in the positive group and the higher scored $K$ patches in the negative group for training.
    ii. Max-Min: select the higher scored $K$ patches in the positive group and the lower scored $K$ patches in the negative group for training.
    iii. Random: in the positive group, select patches with a probability greater than 0.9 and the higher scored $K$ patches, take the union of them, and randomly select $k$ patches; in the negative group, randomly select $k$ from all patches for training.
3. These strategies allowed to define an upper bound of space (Max-Max criterion) and a lower bound of space (Max-Min criterion), while Random lies between the upper and lower bounds. Max-Max and Max-Min criteria were used to select patches from the original training set to build a new supervised dataset that was later used to retrain both networks with the Random criterion to get the final model.

At inference time, the trained models are applied in parallel and the output combined to access the final WSI prediction.

## 13 – Biocenas

The team Biocenas based its method on the work of Ilse and colleagues [10]. The model consisted of a feature extractor with two convolutional-ReLU-max-pooling stages, with 3x3 and 4x4 kernels, followed by two fully connected-ReLU-dropout stages, with 512 hidden neurons each. The resulting feature vector was then filtered by an attention network with a fully connected layer with 512 input neurons and 256 output neurons, with a hyperbolic tangent activation and another fully connected layer with 1 output neuron that selects the most relevant patch. A final fully connected layer with an input size of 512 and sigmoid activation function classifies the representative patch.

Input images were pre-processed by thresholding (Otsu threshold) a downsampled version of the WSI (ratio 1:32) and tiling the tissue regions (downsampled at ratio 1:16) into 32x32 pixel patches. During training, tiles from a single WSI were aggregated in bags of different lengths (randomly chosen without repetition from a Gaussian distribution of $\mu = 50$ and $\sigma^2 = 10$) with a label equal to the label of the original WSI.

## 14 – HEROH

Two residual networks [9], pre-trained on ImageNet [8] were used to build the model of team HEROH. The first, a ResNet18 was trained to classify tissue patches of size 128x128 pixel into one of six classes: tumor, normal, stroma, fat, necrosis, and immune cells. Patches classified as tumor with a probability above 90% were kept and the remaining discarded. To train the first network, a manually annotated dataset was created from 15 WSI, by tiling and standard data augmentations. The patches classified as tumor were then color-deconvolved to extract the hematoxylin channel and an Otsu threshold was applied to identify nuclei. A ResNet50 was then trained to predict the HER2 status of each patch. Each WSI was then classified according to the number of positive and negative patches and the corresponding probabilities. The probabilities assigned to all patches classified as positive were added and stored in a variable $p$, while the probabilities assigned to all patches classified as negative were added and stored in a variable $n$. The WSI was classified as positive if and only if $p/(p+n) > 0.33$.

## 15 – Reza Mohebbian

The team Reza Mohebbian used a model based on user-defined features. After down-scaling the WSI by a factor of 4, each image is converted into 64 grids and processed for feature selection from each grid. A manually selected threshold was considered and all pixels with intensities above 50 were discarded and an active contour algorithm [11] was then applied to identify cells and create a cell mask. From the cell mask, the centroids of each cell were determined. In parallel, the KD-tree algorithm was used to evaluate the density and displacements of cells in each region, and the gray-level co-occurrence matrix was calculated to access features such as contrast, correlation, energy, and homogeneity. Finally, all the features were concatenated and a Particle Swarm Optimization algorithm [12] was used to estimate the weights of features to allow predictions.

## 16 – mindmork

The team mindmork developed a model consisting of three stages. In the first stage, a sliding window is used to tile the WSI into 256 by 256 pixel patches and a K-Means algorithm [13, 14] with 2 clusters is applied to identify tissue regions. Patches containing tissue are, in the second stage, fed to a U-Net [15] model that generates an 8-channel output with information on lymphocytes, fibroblasts, tumor nuclei, collagen fibers, and inter-nuclear space. Tiles predicted by the U-Net are then subjected to feature extraction and 20 features, including nuclei density, collagen fiber length, endpoint density, fiber orientation, and texture quantifications, are exported per tile and aggregated per WSI. In the third stage, an Xception model [16] is used to classify the WSI.

## 17 – Institute of Pathology Graz

The team Institute of Pathology Graz developed a model based on standard image analysis techniques to extract features of interest. A QuPath [17] script was developed consisting of a color deconvolution protocol followed by a nuclei segmentation (QuPath function WatershedCellDetection). For each nucleus, the parameters described in Table S1 were computed and aggregated into a feature database. A custom convolutional neural network (CNN) was then created in the generated database to classify each WSI based on the described features.

Table S1: Parameters quantified per nucleus. The groups for the nucleus area were defined as small, if $area < 18$; medium if $18 \leq area < 54$; large if $54 \leq area \leq 82$; or extra-large if $area > 82$. The circularity is considered small if it is equal to or smaller than 0.5, medium if it is greater than 0.5 and smaller than 1, and large if it is equal to 1.

| PARAMETER GROUP | PARAMETER |
| --- | --- |
| AREA | mean nucleus area |
| | number of nuclei with a small area |
| | number of nuclei with a medium area |
| | number of nuclei large area |
| | number of nuclei with extra-large area |
| CIRCULARITY | mean nucleus circularity |
| | mean nucleus circularity in group small |
| | mean nucleus circularity in group medium |
| | mean nucleus circularity in group large |
| | mean nucleus circularity in group extra large |
| HEMATOXYLIN OPTICAL DENSITY (OD) | mean nucleus hematoxylin od per nucleus |
| | mean nucleus hematoxylin od in group small |
| | mean nucleus hematoxylin od in group medium |
| | mean nucleus hematoxylin od in group large |
| | mean nucleus hematoxylin od in group extra-large |
| AREA CIRCULARITY RATIO | mean nucleus area/circularity |
| | mean nucleus area/circularity in group small |
| | mean nucleus area/circularity in group medium |
| | mean nucleus area/circularity in group large |
| | mean nucleus area/circularity in group extra-large |
| AREA CIRCULARITY DENSITY RATIO | mean nucleus area/circularity/hematoxylin od mean in group small |
| | mean nucleus area/circularity/hematoxylin od mean in group medium |
| | mean nucleus area/circularity/hematoxylin od mean in group large |
| | mean nucleus area/circularity/hematoxylin od mean in group extra-large |
| AREA COUNTS | number of nuclei with small circularity |
| | number of nuclei with medium circularity |
| | number of nuclei with large circularity |

## 18 – katherandco

Team katherandco manually annotated tumor tissue on each WSI using QuPath [17]. Patches with size 512x512 pixels were then created and only the regions containing invasive breast cancer were kept for the next steps. Data augmentation was applied and a ResNet50 [4] pre-trained on the ImageNet dataset [8] was then re-trained on the resulting dataset. During the classification step, each WSI is classified according to the score achieved by its highest scored patch.

## 19 – QUILL

The team QUILL used a model based on a DenseNet architecture [7]. After splitting the WSI on the training dataset into 121828 superpixel patches labeled according to the WSI they were cropped from, a DenseNet network was trained and used as a feature extractor. The mean shift clustering algorithm [18] was then used to identify the modal vector in the feature space, and the feature vectors more distant (one positive and another negative) from the modal are considered as representative and used to predict the WSI status.

## 20 – HEROHE_Challenge

Team HEROHE_Challenge built a custom convolutional neural network and pre-trained it on the Cfar10 images dataset [19]. Then, each WSI was tiled into 200x200 pixels patches with 10 pixels overlap to extract a feature vector per patch. A K-Means algorithm [13, 14] was then used to group the feature vectors into 50 classes and for each class, the vector closer to the center of the class was selected as representative of its class. Finally, the representative feature vectors were concatenated and the XGBoost method [20] was used to distinguish between HER2 positive and negative slides.

## 21 – UC-CSSE

The team UC-CSSE team used an Xception network [16] pre-trained on the CAMELYON16 dataset [21] to identify tumor regions. Nuclei were then segmented with an ensemble of networks adapted from the model (https://github.com/selimsef/dsb2018_topcoders) that won the 2018 Data Science Bowl (https://www.kaggle.com/c/data-science-bowl-2018). The ensemble consisted in the use of 3 convolutional neural networks: DenseNet169 [7], ResNet34, and ResNet101 [4]. After nuclei segmentation, a hand-crafted feature vector was computed, including shape, texture, and neighborhood statistic features. An ensemble of random forest [22], extra trees [23], and gradient boosting [24] classifiers were finally developed to classify each WSI according to its HER2 status.

## References

1. Szegedy, C., et al. Rethinking the inception architecture for computer vision. in Proceedings of the IEEE conference on computer vision and pattern recognition. 2016.
2. Oner, M.U., H.K. Lee, and W.-K. Sung. Weakly Supervised Clustering by Exploiting Unique Class Count. in International Conference on Learning Representations. 2020. https://openreview.net/forum?id=B1xIj3VYvr.
3. Campanella, G., et al., Clinical-grade computational pathology using weakly supervised deep learning on whole slide images. Nat Med, 2019. 25(8): p. 1301-1309.
4. He, K., et al. Deep Residual Learning for Image Recognition. in 2016 IEEE Conference on Computer Vision and Pattern Recognition (CVPR). 2016.
5. Dietterich, T.G., R.H. Lathrop, and T. Lozano-Pérez, Solving the multiple instance problem with axis-parallel rectangles. Artificial Intelligence, 1997. 89(1): p. 31-71.
6. Hu, J., L. Shen, and G. Sun. Squeeze-and-excitation networks. in Proceedings of the IEEE conference on computer vision and pattern recognition. 2018.
7. Huang, G., et al. Densely connected convolutional networks. in Proceedings of the IEEE conference on computer vision and pattern recognition. 2017.
8. Deng, J., et al. ImageNet: A large-scale hierarchical image database. in 2009 IEEE Conference on Computer Vision and Pattern Recognition. 2009.
9. Xie, S., et al. Aggregated residual transformations for deep neural networks. in Proceedings of the IEEE conference on computer vision and pattern recognition. 2017.
10. Ilse, M., J. Tomczak, and M. Welling. Attention-based deep multiple instance learning. in International conference on machine learning. 2018. PMLR.
11. Chan, T.F. and L.A. Vese, Active contours without edges. IEEE Transactions on Image Processing, 2001. 10(2): p. 266-277.
12. Kennedy, J. and R. Eberhart. Particle swarm optimization. in Proceedings of ICNN'95 - International Conference on Neural Networks. 1995.
13. Lloyd, S., Least square quantization in PCM. Bell Telephone Laboratories Paper. Published in journal much later: Lloyd, SP: Least squares quantization in PCM. IEEE Trans. Inform. Theor.(1957/1982), 1957. 18.
14. MacQueen, J. Some methods for classification and analysis of multivariate observations. in Proceedings of the fifth Berkeley symposium on mathematical statistics and probability. 1967. Oakland, CA, USA.
15. Ronneberger, O., P. Fischer, and T. Brox. U-Net: Convolutional Networks for Biomedical Image Segmentation. 2015. Cham: Springer International Publishing.
16. Chollet, F. Xception: Deep learning with depthwise separable convolutions. in Proceedings of the IEEE conference on computer vision and pattern recognition. 2017.
17. Bankhead, P., et al., QuPath: Open source software for digital pathology image analysis. Sci Rep, 2017. 7(1): p. 16878.
18. Comaniciu, D. and P. Meer, Mean shift: a robust approach toward feature space analysis. IEEE Transactions on Pattern Analysis and Machine Intelligence, 2002. 24(5): p. 603-619.
19. Krizhevsky, A., Learning Multiple Layers of Features from Tiny Images. University of Toronto, 2012.
20. Chen, T. and C. Guestrin, XGBoost: A Scalable Tree Boosting System, in Proceedings of the 22nd ACM SIGKDD International Conference on Knowledge Discovery and Data Mining. 2016, Association for Computing Machinery: San Francisco, California, USA. p. 785–794.
21. Ehteshami Bejnordi, B., et al., Diagnostic Assessment of Deep Learning Algorithms for Detection of Lymph Node Metastases in Women With Breast Cancer. JAMA, 2017. 318(22): p. 2199-2210.
22. Breiman, L., Random Forests. Machine Learning, 2001. 45(1): p. 5-32.
23. Geurts, P., D. Ernst, and L. Wehenkel, Extremely randomized trees. Machine Learning, 2006. 63(1): p. 3-42.
24. Friedman, J., T. Hastie, and R. Tibshirani, The elements of statistical learning. Vol. 1. 2001: Springer series in statistics New York.